\documentclass[aps.pra,onecolumn,showpacs,preprintnumbers,amsmath,amssymb]{revtex4}
\usepackage{mathrsfs}
\usepackage{graphicx}
\usepackage{amsmath}

\begin{document}

\title{Dynamical Lamb Effect in a Tunable Superconducting Qubit-Cavity System}
\author{D. S. Shapiro$^{1,2}$, A. A. Zhukov$^{1,3}$, W. V. Pogosov$^{1,4}$, Yu. E. Lozovik$^{1,5,6}$}

\affiliation{$^1$Center for Fundamental and Applied Research, N. L. Dukhov All-Russia Research Institute of Automatics, 127055 Moscow, Russia}
\affiliation{$^2$V. A. Kotel'nikov Institute of Radio Engineering and Electronics, Russian Academy of Sciences, 125009 Moscow, Russia}
\affiliation{$^3$National Research Nuclear University (MEPhI), 115409 Moscow, Russia}
\affiliation{$^4$Institute for Theoretical and Applied Electrodynamics, Russian Academy of
Sciences, 125412 Moscow, Russia}
\affiliation{$^5$Institute of Spectroscopy, Russian Academy of Sciences, 142190 Moscow region,
Troitsk, Russia}
\affiliation{$^6$Moscow Institute of Electronics and Mathematics, HSE, 101000 Moscow, Russia}

\begin{abstract}
A natural atom placed into a cavity with time-dependent parameters can be parametrically excited due to the interaction with the quantized photon mode. One of the channels of such a process is the dynamical Lamb effect, induced by a nonadiabatic modulation of atomic level Lamb shift.
However, in experiments
with natural atoms it is quite difficult to isolate this effect from other mechanisms
of atom excitation. We point out that a transmission line cavity coupled with a superconducting
qubit (artificial macroscopic atom) provides a unique platform for the observation of the dynamical Lamb effect. A key idea is to exploit a dynamically tunable qubit-resonator coupling, which was implemented quite recently. By varying nonadiabatically the coupling, it
is possible to parametrically excite a qubit through a nonadiabatic modulation of the Lamb shift, even if the cavity was initially empty.
A dynamics of such a coupled system is studied within the Rabi model with time-dependent coupling constant and beyond the rotating wave approximation. An
efficient method to increase the effect through the
periodic and nonadiabatic switching of a qubit-resonator coupling energy is proposed.
\end{abstract}

\author{}
\maketitle
\date{\today }

\section{Introduction}

Superconducting circuits with Josephson junctions can be used for
quantum computing as qubits \cite{Nation, YouNori, Blais, mooij,
shs, Manko}. These systems are macroscopically large, but they demonstrate
quantum behaviour, which allows them to be treated as artificial
atoms. Superconducting qubits can be integrated with microwave
waveguides, while the photon field in these waveguides is quantized.
Such systems give rise to the effects of Rabi oscillations
\cite{Vijay, Fink, Bertet, Wallraff}, quantum feedback \cite{sdz},
GHz photons emission \cite{Romero}, while many-qubit systems form
subwavelength quantum metamaterials \cite{Rakhmanov, Astafiev,
Ustinov,Shapiro}. A well known Rabi model \cite{Rabi} is applicable for the
description of qubit-photon quantum system \cite{Nation}.

Moreover, superconducting circuits integrated with microwave
resonators provide a unique platform for an observation of cavity
quantum electrodynamical (QED) nonstationary phenomena which can
hardly be studied in more traditional experiments. This happens because
(i) resonator frequencies in superconducting systems are nearly six orders
of magnitude lower than frequencies in optical systems, (ii) superconducting
qubit-cavity systems demonstrate high and fast tunability of their main parameters.

One of the examples of such nonstationary QED phenomena
is the dynamical Casimir effect, predicted long time ago
\cite{Moore}. According to the initial idea, an 'empty' space
between two mirrors can emit photons due to vacuum fluctuations,
provided these mirrors are rapidly moving with respect to each
other. In order to observe this effect, one has to move mirrors with
the speed approaching the speed of light, which seems to be
unrealistic for a direct experimental realization. For this reason,
there were several other proposals how the dynamical Casimir effect
can be observed, see, e.g., Refs.
\cite{Yablonovich,Lozovik-plasma,Dodonov}. In all these schemes, it
was suggested that an effective boundary condition at 'mirrors'
could be varied nonadiabatically, instead of moving massive 'mirrors'
themselves. It is remarkable that it was a superconducting
system, which led to the first observation of the dynamical Casimir
effect \cite{DCE1,DCE2}. This was achieved by using an additional
SQUID at the end of the waveguide in order to tune an effective
boundary condition by varying in time the magnetic flux through the
SQUID.

A presence of an additional atom in a cavity with time-dependent
parameters leads to other nonstationary QED phenomena, since an atom
and a photon field can interact with each other
\cite{Heinzen,Pokrovski,Lozovik-lett,Lozovik1,review}. For instance,
an atom can be parametrically excited, even if the cavity was
initially empty. It was shown in Ref. \cite{Lozovik1} that there are
several channels for such a process.  In the case of a
nonadiabatical modulation of cavity parameters, there are two
channels of such a process. The most obvious mechanism is due to the
absorbtion of Casimir photons. Another channel is due to the
nonadiabatic modulation of atomic Lamb shift: virtual states of
atom-photon coupled system are transformed into real states. This
phenomenon can be called a dynamical Lamb effect. Note that the
existence of a static Lamb shift for superconducting qubit energy
levels was verified experimentally in Ref. \cite{Lamb}. In contrast
to natural atoms, the effect can be significantly enhanced, since a
strong coupling regime between the artificial macroscopic atom and
the resonator is achievable.

In practice, it is quite difficult to isolate the mechanism of atom
excitation due to the dynamical Lamb effect from another channel due
to the the absorbtion of Casimir photons. In order to overcome this
difficulty, it was suggested in Ref. \cite{Lozovik1} that a single
natural atom can be passed through the resonator consisting of two
cylindrical cameras of different diameters which are characterized
by two unequal Lamb shifts. When leaving one camera and entering
another one, the dynamical Lamb effect can occur, so that an atom
can be parametrically excited. A specific feature of this scheme is
that, in this case, no Casimir photon appears, so that the dynamical
Lamb effect can be, in principle, isolated. Unfortunately, a direct
experimental implementation of this idea is rather difficult.

The main goal of this paper is to suggest a superconducting
qubit-resonator system as a suitable platform for the observation of
the dynamical Lamb effect instead of a natural atom-cavity system.
An astonishing property of superconducting qubits is a high
tunability of their parameters \textit{in situ} during an experiment
\cite{Devoret,Girvin}. For instance, one can modulate qubit
excitation energy to manipulate qubits by using this degree of
freedom \cite{PRB2013}. Moreover, due to a recent technological
progress, it becomes realistic to vary dynamically not only the
resonator frequency,
 but also a coupling between
the artificial atom (transmon qubit) and photon field with the
frequency of a resonator or even faster. This quantity already can
be tuned in sufficiently wide ranges by an auxiliary SQUID, see,
e.g., Ref. \cite{tunable_qubit} and references therein. Using this
additional degree of freedom, one can mimic a single atom passing
from one camera to another one by a single switching of a qubit
between the two resonators. Namely, this scheme can be mapped on a
superconducting qubit at rest integrated with the resonator A, while
at certain moment it is switched nonadiabatically to the initially
empty resonator B. It is expected that this switching can be
accompanied by the parametric excitation of a qubit. In this scheme,
a channel of a qubit excitation through the absorbtion of real
Casimir photons is suppressed, because in the weak coupling regime
the perturbation of the cavity eigenmodes is negligible. Thus,
instead of relying on a physical motion of an atom, one can simply
tune a magnetic flux through a special auxiliary SQUID.

Actually, a high tunability of superconducting circuit systems
allows even for a significant simplification of this mapping.
Instead of utilizing two resonators, one can use a single one.
Indeed, one can consider a qubit initially uncoupled from the
resonator, while at certain moment this coupling is nonadiabatically
switched on leading again to a possibility of a qubit excitation and
photon generation. Furthermore, a superconducting qubit-resonator
system can be made in a strong coupling regime which must increase
dramatically the probability of qubit excitation due to the
dynamical Lamb effect. Note that an important condition to observe
the dynamical Lamb effect is also that all switchings have to be
performed nonadiabatically. This implies that the typical switching
time to be smaller than the resonator frequency. Such a regime
becomes possible due to the very recent progress
\cite{tunable_qubit}.

In order to describe theoretically the dynamical Lamb effect in the system with the tunable qubit-photon coupling, we use the Rabi model.
To treat the dynamical Lamb effect, we have to go beyond
the rotating wave approximation \cite{JCmodel} (RWA) in order to take into account counter-rotating processes.
These are the photon creation with the simultaneous qubit excitation and the opposite process. Such processes in superconducting qubit-resonator systems are not at all illusive, since they show up in experiments with strong-coupled systems, see, e.g., Ref. \cite{vNature}.
We take into account counter-rotating wave terms by constructing
a perturbation theory around the stationary RWA solution. In
addition, we solve a time-dependent Schr\"{o}dinger equation
numerically. We find that our simple analytical formula for the
qubit excitation probability gives excellent results in the case,
when the interaction constant is switched nonadiabatically only
once.

Our treatment does not only yield the description of the system's dynamics but also allows us to suggest a special trick which can be used to enhance a dynamical Lamb effect. Namely, we show that the qubit excitation probability can be dramatically
increased provided that the coupling constant is switched on and off
periodically with the period twice larger than the resonator
frequency (each switching is nonadiabatical). Previously, similar periodic drivings have been proposed to enhance other cavity QED effects in a context of superconducting circuits, such as the dynamical Casimir effect \cite{DCE1,DCE2} or Lamb shift \cite{Lambshift}. For such a periodic driving, our analytical perturbative result yields
qualitatively correct description of the system's dynamics. We also
examine the statistics of photon states, generated due to the coupling
constant dynamics. We show that a significant squeezing of these
states can be achieved. This result can be of practical importance.

We wish to stress that the main focus of this paper is not at all to develop a
new method for superconducting qubit manipulation, but to suggest a scheme for the experimental
realization of the dynamical Lamb effect. Note also that from the viewpoint of a quantum computing, the dynamical
Lamb effect can be either positive or parasitic depending on a
particular situation. For instance, it can lead to the undesirable
qubit excitation. Therefore, the understanding and
control of the dynamical Lamb effect is of importance both from the
viewpoint of a fundamental physics and possible applications.

This paper is organized as follows. In Section II, we present  our
idea and outline a theoretical model used. In Section III, we find
qubit excitation probability for different regimes of qubit-resonator
coupling dynamics. In Section IV, we address correlation
functions for both the photon field and qubit degrees of freedom
upon coupling energy dynamics. We conclude in Section V.

\section{Model}

There are different types of superconducting qubits. All of them
represent a superconducting ring containing  several Josephson
junctions, but they differ from each other by ratios of their
charging energies and Josephson energies. In order to study
experimentally the dynamical Lamb effect, we need a qubit which
allows for the dynamical tunability between its own degrees of freedom and
photon field with the frequency of a resonator or even faster.
Actually, these are charge qubits (transmons) and flux qubits
that seems to be more suitable for such a task \cite{tunable_qubit}.

The cavity is a coplanar waveguide where the photon mode frequency
$\omega$ is large compared to the characteristic energy of
interaction between the qubit and photon degrees of freedom (weak
coupling).  This limit supposes that the resonator is operated in
the single mode regime. In practice, in order to enhance the
dynamical Lamb effect, a strong coupling regime might be more
appropriate. However, it is much more difficult to address this
regime within the analytical treatment. We, therefore, focus on a
weak coupling regime and show that even in this case a significant
effect is possible.

The inductive coupling between a qubit and resonator photons is achieved
by an auxiliary SQUID operating at the same GHz frequencies \cite{tunable_qubit}. This
opportunity is of a crucial importance for an observation of the
dynamical Lamb effect. The coupling energy can be changed either by a
single switching or via a periodic modulation. The latter approach
will be shown to be much more efficient.

The qubit-resonator system can be described in terms of the Rabi
model \cite{Rabi,JCmodel}, which is widely used in quantum optics. The
total Hamiltonian of this model takes into account photons at the
single mode  $\omega$ and the qubit with the bare excitation frequency
$\epsilon$ with the coupling $V$ between them
\begin{equation}
H=\omega a^{\dagger }a + \frac{1}{2} \epsilon (1+\sigma_{3})+V,
\label{Hamiltonian}
\end{equation}
where $a^{\dagger }$ and $a$ are secondary quantized boson operators
of the photon field,  while Pauli matrices
$\sigma_{3}=2\sigma_{+}\sigma_{-}-1$, $\sigma_{+}$, $\sigma_{-}$ act
in the space of qubit states. The operator $V$ reads as
\begin{equation}
V=g(a+a^{\dagger })(\sigma_{-}+\sigma_{+}),
\label{FullV}
\end{equation}
where $(a+a^{\dagger })$ and $\sigma_{-}+\sigma_{+}$, up to
numerical factors, are nothing but the electric field and dipole
moment operators, respectively, while $g \ll \omega$ is a coupling
constant.Under this assumption, a perturbation of the cavity eigenmodes
is negligible, so that a channel of qubit excitation via the absorbtion of
real Casimir photons is not relevant.

A decoherence $\kappa$ and relaxation rate $\Gamma$ are of
importance  for real qubits, available in modern experiments.
Typically, these two quantities turn out to be larger or of the same
order as $g$. However, nowadays there is a steady and rather fast
progress in decreasing $\kappa$, $\Gamma$, so we will concentrate on
the limit $g\gg \kappa,\Gamma$.
% We will also briefly discuss the regime $g \sim \kappa,\Gamma$.

In the case of a stationary system, a perturbation theory in
operator $V$ can be developed rather easily. It was shown in Ref.
\cite{Lozovik1} that $V$ produces contributions to the
eigenenergies, which, in leading order, are independent on a photon level
number but depend only on atomic level number.
These contributions are interpreted as a Lamb shift, since
these are virtual photons which are responsible for the effect. The
approach of Ref. \cite{Lozovik1} is justified in the off-resonant
regime $|\Delta|\gg g$, where $\Delta=\epsilon-\omega$ is a detuning
frequency. It predicts the following shift for the atomic ground
state energy:
\begin{equation}
 \epsilon_{L}=-\frac{g^2}{\omega+\epsilon}. \label{lamb_off-res}
 \end{equation}

 Actually, there is another approach to the problem, which has some advantages.
 Namely, the
  operator $V$ can be split into the sum of two terms,
 \begin{equation}
 V=V_1+V_2,
  \label{split}
\end{equation}
where
 \begin{equation}
V_1= g(a \sigma_{+} + a^{\dagger } \sigma_{-})
\label{V1}
\end{equation}
is the rotating wave contribution, conserving the excitation number, and
 \begin{equation}
V_2=g(a^{\dagger } \sigma_{+} + a \sigma_{-})
\label{V2}
\end{equation}
is the counter-rotating term, which breaks this symmetry, but conserves a parity.

We will focus on the situation, when the qubit-cavity system is near
the resonance, $\epsilon \simeq \omega$, since in this case the
dynamical Lamb effect is strongest.  This resonant regime can be
achieved by tuning qubit frequency  $\epsilon$ \textit{in situ} via
an  external  magnetic flux, threading the qubit loop.
Near the resonance and if $g$ is also independent on time, the fast
oscillating term $V_2$ is usually omitted, while the resulting
Hamiltonian is known to be exactly integrable. This well known
solution is outlined in Appendix A. We have to take $V_2$ into account since
it is precisely this operator, which is responsible for the stationary
Lamb shift and, hence, for the dynamical Lamb effect as well. In the
stationary case, $V_2$ can be treated as a perturbation. This
approach is also utilized in Appendix A for the evaluation of the
stationary Lamb shift in a resonant regime. Moreover, it will be
demonstrated that a similar splitting of the Hamiltonian gives
excellent results in certain nonstationary situations as well.

\section{Dynamical Lamb effect}

\subsection{Qualitative picture}

This section deals with the main subject of our paper. Namely, we
consider a coupled qubit-resonator system, when the coupling
energy $g$ is time dependent. We assume that in the initial
moment, the qubit and the resonator were not coupled, so that the
initial state of the whole system was $|  0 \downarrow\rangle$. When
the coupling is switched on, higher-order photon states start to
play their role, as well as a qubit excited state. In the case of a
full resonance, this process can be illustrated by a scheme,
presented in Fig. 1. Apart from the bare ground state $|  0
\downarrow\rangle$, there are two sets of doubly-degenerate excited
bare states. These states feel each other through the operator $V$,
which is a sum of two contributions. The first one, $V_1$, conserves
the excitation number, i.e., a total number of photons plus
a qubit excitation number (0 or 1). The second one,
$V_2$, changes the excitation number by 2 by adding/removing one
photon and simultaneously switching qubit to the excited/ground
state.

Thus, we see that the bare ground state can be perturbed by $V_2$
only, which leads to the occupation of the $|  1 \uparrow\rangle$
state with nonzero probability. In its turn, this latter state is connected by $V_2$ only
with $|  0 \downarrow\rangle$ and by $V_1$ only with $|  2
\downarrow\rangle$. By continuing this process, we see that (i) only
the states with even excitation number are populated, (ii) the
states with the same excitation number are linked only via $V_1$,
(iii) the states with the excitation numbers differing by 2 are
linked by $V_2$ only.

Let us assume that tunable coupling $g$ is not oscillating too fast.
In this case, $V_2$ must be responsible to fast oscillations, while
$V_1$ corresponds to slow oscillations with frequencies essentially
controlled by the Rabi frequency. Therefore, we may expect that
$V_2$ can be treated perturbatively, although both $V_1$ and $V_2$
are formally proportional to the same parameter $g$ (a small
parameter responsible for this feature in a resonant case is
$g/\omega$). This implies that the qubit excitation probability is
going to experience oscillations with a frequency of the order of
the Rabi frequency. Such oscillations are induced by processes
between the states of the same excitation number. By solving
time-dependent Schr\"{o}dinger equation numerically, we will also
show that this qualitative picture is indeed correct. It is also of
importance to note that the photon-qubit dynamical coupling results
in the occupation of all photon states, i.e., with both odd and even
photon numbers. This feature, which can be probed in experiments, is
in contrast with the dynamical Casimir effect, which leads to the
occupation of the states with only even number of photons
\cite{Lozovik1}.

Notice that slow Rabi-like oscillations
show up even in the first order of the perturbation theory around the RWA solution.
In contrast, when applying a perturbation theory around
bare states, all terms of an infinite expansion in $V$ must be taken into account in order to
recover such oscillations. This is one of the advantages of the theoretical
approach utilized in the present paper.

\begin{figure}[h]
\center\includegraphics[width=0.35\linewidth]{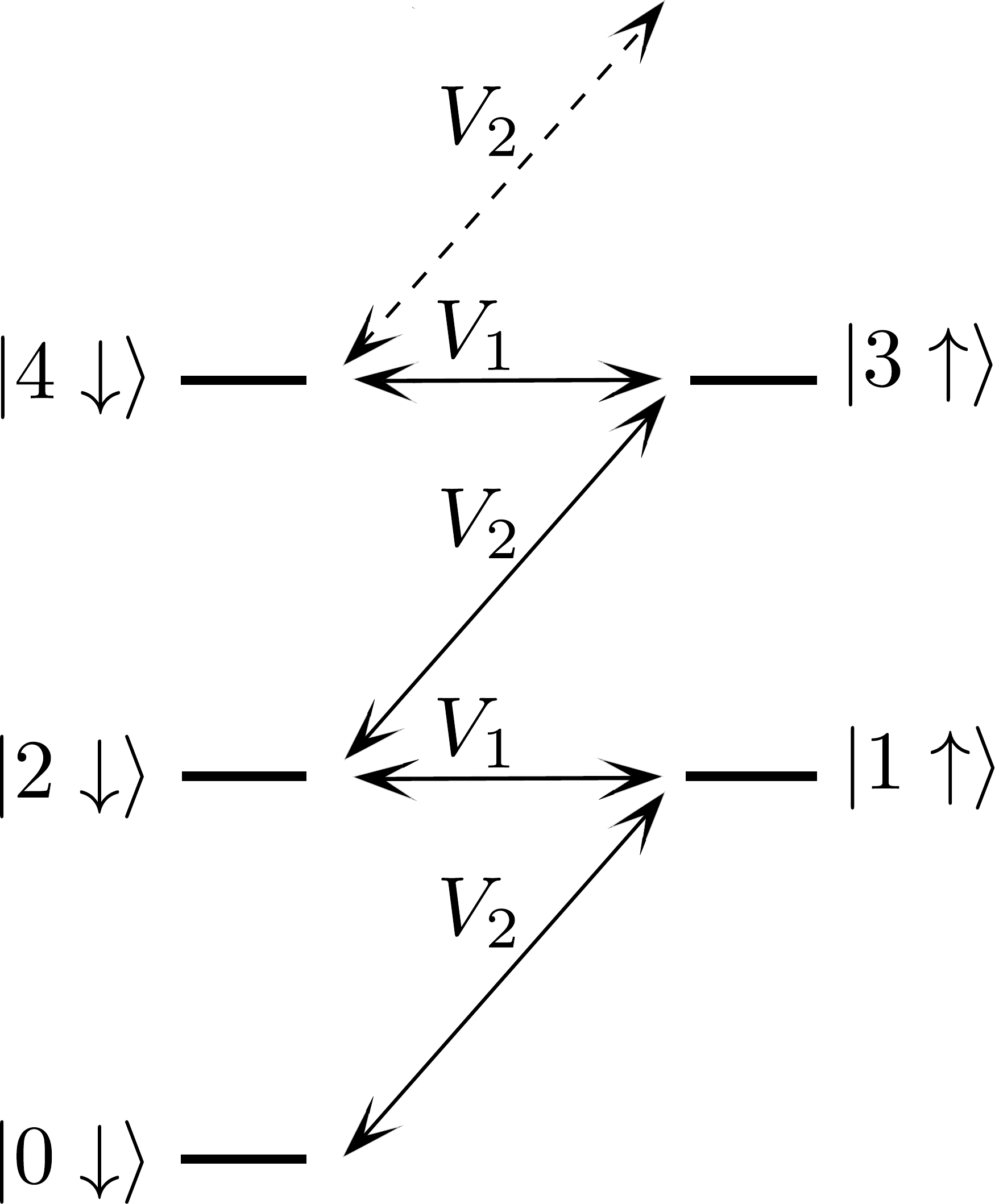}
\caption{ \label{qualitative}A scheme illustrating an internal structure of the full wave function projected into states of the qubit-waveguide system without coupling (see in the text).}
\end{figure}

\subsection{Time-dependent perturbation theory}

As a zero-order approximation, we take the \emph{stationary} RWA
Hamiltonian,  in which the coupling constant is equal to the time
average value $\bar g=\langle g(t)\rangle_t$ (averaging is performed
after the coupling is switched on) and remaining
\emph{nonstationary} terms are treated as a perturbation. This
perturbation contains counter rotating wave operator $V_2$ and a
rotating wave contribution, which is due to deviations of $g$ from its time
averaged value. Namely, the full Hamiltonian is represented as
\begin{equation}
H(t)=H_{RWA}+V(t),
\end{equation}
where
\begin{eqnarray}
H_{RWA}=\omega a^{\dagger }a+ \frac{1}{2} \epsilon (1+\sigma_{3}) +\bar g (a \sigma_{+} + a^{\dagger } \sigma_{-}), \notag \\
V(t)=g(t)(a^{\dagger }\sigma_{+} + a\sigma_{-}) +(g(t)-\bar g)(a\sigma_{+} + a^{\dagger }\sigma_{-}).
\label{spli}
\end{eqnarray}

Let us now consider the evolution of the wave function
$\psi(t=0) =|0\downarrow\rangle$  assuming that a coupling is
switched on at $t=0$. Then, at the moment $t>0$, the wave function is
given by the standard expression
 \begin{equation}
 \psi(t)= e^{-iH_{RWA}t} \mathcal{T } \exp\left( -i \int\limits_0^t e^{iH_{RWA}\tau} V(\tau) e^{-iH_{RWA}\tau} d\tau \right) \psi(0), \label{solution}
 \end{equation}
 where $\mathcal{T}$ stands for the time ordering operation. In our
 calculations we take into account only first order terms in $\mathcal{T }$-exponent
 series expansion in Eq. (\ref{solution}). This approximation is further justified by a direct
 comparison with the result of the solution of the time-dependent Schr\"{o}dinger equation.

 \subsection{Driving by a single switching of a coupling constant}
 \label{lamb_dynamic_single}

 We now study the evolution of the wave function in a full
 resonance regime,  $\epsilon=\omega$, after the sudden switch of the coupling $g$,
 which then stays constant in time. Thus, we have $g(t)=g\theta(t)$ and $\bar g=g$, so that
 the second contribution in operator $V(t)$, as given by Eq. (\ref{spli}), vanishes.
  Using Eq. (\ref{solution}), we obtain
 \begin{eqnarray}
  \psi(t)=|0 \downarrow\rangle - \frac{g}{2\omega}(e^{-2i\omega t}\cos \sqrt{2}gt-1)|1 \uparrow\rangle-i\frac{g}{2\omega}e^{-2i\omega t}\sin \sqrt{2}gt|2 \downarrow\rangle .
  \label{single_switching}
 \end{eqnarray}
 In this approximation, the qubit excitation probability as a result of the instantaneous  switching, is given by
 \begin{equation}
w_\uparrow(t)=\frac{g^2}{8\omega^2}\left(3+\cos 2\sqrt{2}gt -4\cos 2\omega t\cos \sqrt{2}gt\right) . \label{prob_single_switching}
 \end{equation}
This probability corresponds to the dynamical Lamb effect and includes both slow
and fast oscillations. In the superconducting systems with weak
coupling between photon field and qubit, the probability is as small
as $g^2/ \omega^2$. Note that the amplitude of this quantity is proportional to the square of the change of the static Lamb shift, in accordance with Ref. \cite{Lozovik1}.

The
number of generated photons as a function of time $n_{ph}(t)=\langle
\psi(t)| a^{\dagger }a|\psi(t) \rangle$ can be also
readily extracted from Eq. (\ref{single_switching}).
We would like to stress that these photons appear as a result of the dynamical Lamb effect.
Our results
for $w_\uparrow(t)$  and $n_{ph}(t)$ at $\omega/g=20$ are plotted in Fig. 2
for $0<t<1.2 T_R$, where the Rabi period is defined as $T_R=\pi/g$.

In order to justify these results, we also solved Schr\"{o}dinger
equation  numerically in the weak coupling regime. The basis was
truncated by typically 10 photon states. An accuracy was verified by
increasing the number of such states and comparison with results for
a smaller basis. The results of such numerical computations show
excellent agreement with our perturbative results for both $w_\uparrow(t)$
and  $n_{ph}(t)$. The relative discrepancy is essentially within few percent.
Such an excellent
agreement justifies our qualitative understanding of the system
dynamics, discussed above.

%both quantities are very small!

\begin{figure}[h]
\center\includegraphics[width=0.5\linewidth]{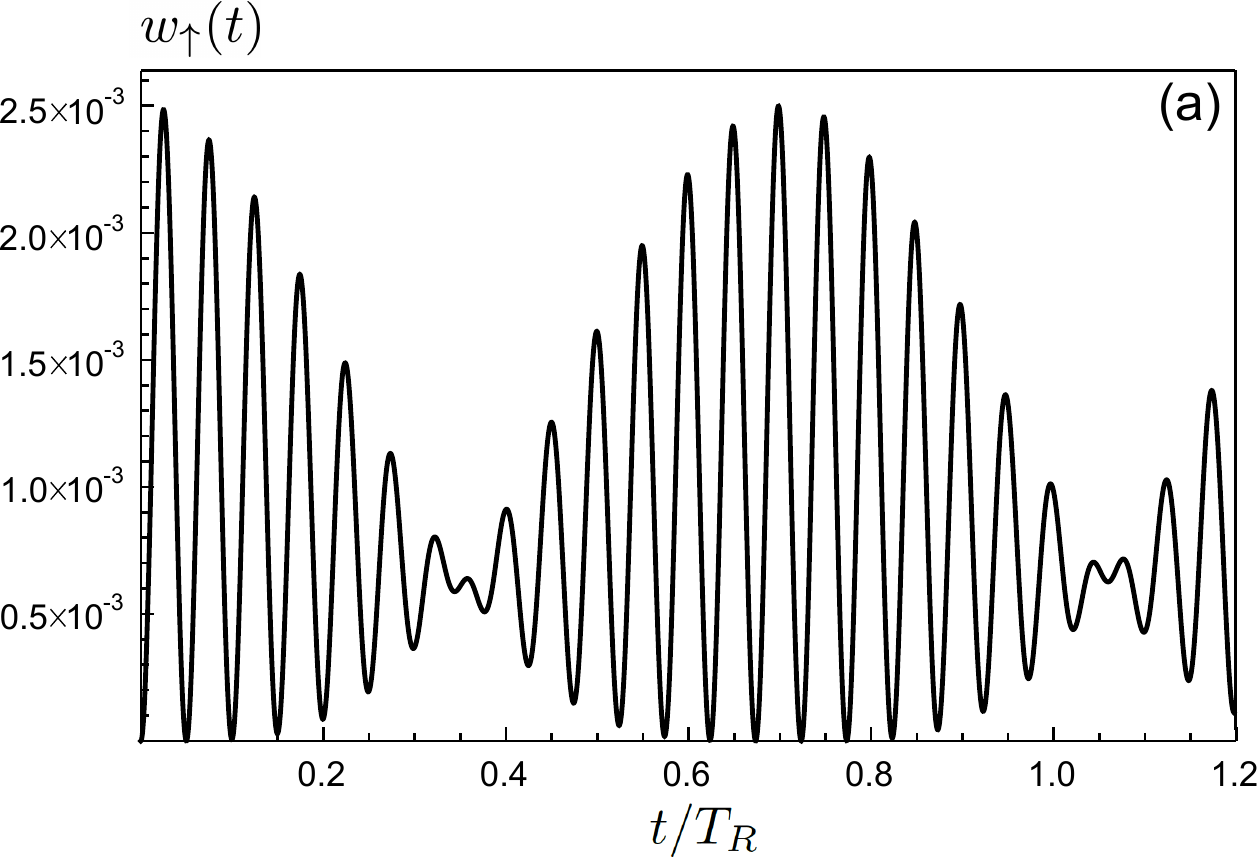}
\center\includegraphics[width=0.5\linewidth]{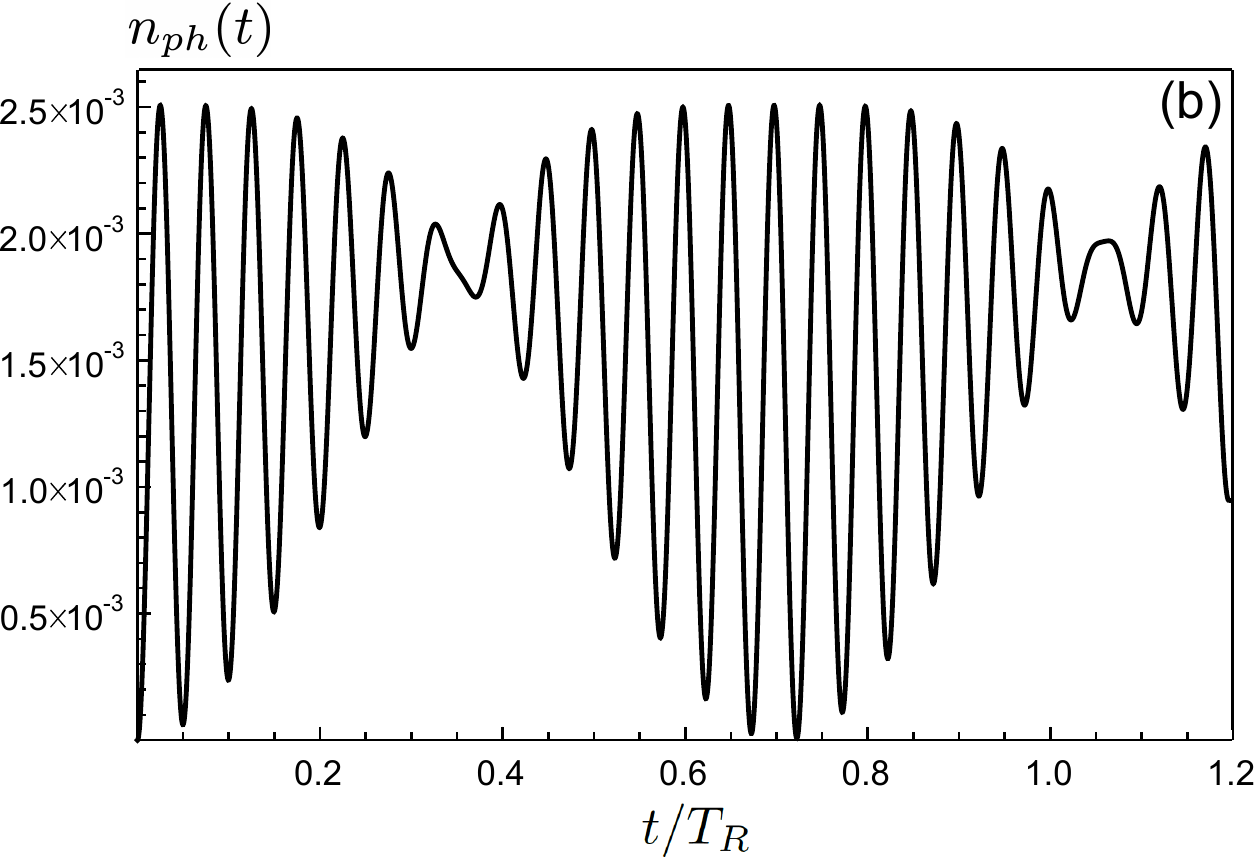}
\caption{ \label{parameteric_num_res} The qubit excitation probability $w_\uparrow(t)$ (a) and the
photon number $n_{ph}(t)$ (b) as functions of time after a single switching of the qubit-waveguide coupling constant at $\omega/g=20$.}
\end{figure}

\subsection{Parametric driving}
 \label{lamb_dynamic_periodic}

Single switching of the qubit-resonator coupling results in the low
probability for the qubit excitation. In order to enhance it, one
can use  more sophisticated modulations of $g$ as a function of time.
We, however, should keep in mind that the dynamical Lamb effect can be isolated only in the case, when this coupling changes nonadiabatically, so that the only one way is to switch it off and on instantaneously.

The excitation probability can be dramatically increased at the
regime of a parametric  resonance, when the driving is induced by
high frequency periodic switching of the qubit-resonator coupling.
%WHY?
In particular, we suggest applying $2\omega$-periodical switching on
and off,  namely,  $g(t)=g\theta(\cos 2\omega t)$. Hence, we take
$\bar g =g/2$ in the Hamiltonian splitting, as given by Eq. (\ref{spli}).
In order to
evaluate qubit excitation probability, we again limit ourselves to
the first order approximation in Eq. (\ref{solution}).

We again assume that $\omega\gg  g$. Therefore, fast
 oscillating terms of the form $e^{i\omega t} g(t)$ can be replaced by
their average values $\langle e^{i\omega t} g(t)\rangle_t $, when
performing integration in Eq. (\ref{solution}), while corrections to
this approximation are negligible as $g/\omega$. Under this
replacement, we find
\begin{eqnarray}
  \psi(t) \simeq  \,|0 \downarrow\rangle + \,\frac{2}{\pi}e^{-2i\omega t}\left(\left(1-\cos \frac{g t}{\sqrt{2}}\right)|2 \downarrow\rangle - i \sin \frac{g t}{\sqrt{2}}\; |1 \uparrow\rangle\right).
 \label{periodic_switching}
 \end{eqnarray}
The corresponding excitation probability is
 \begin{equation}
w_\uparrow(t) \simeq \frac{2}{\pi^2}(1-\cos\sqrt{2} gt). \label{prob_periodic_switching}
 \end{equation}
This  result  describes an approximate Rabi-like oscillatory
behavior of the qubit excitation probability via the dynamical Lamb
effect in the regime of a parametric resonance. Note that
frequences of oscillations of qubit excitation probability and mean
photon number differ by a factor of 2, according to Eq.
(\ref{periodic_switching}). Regardless of the relation between
amplitude of the switching $g$ and $\omega$, the maximum probability
reaches the finite value $4/\pi^2$ on a characteristic time scale
$\sim 1/g$, which depends on the amplitude of the inverse coupling strength
only. In contrast to the probability in the regime of a single
switching (\ref{prob_single_switching}), the time dependence of
$w_\uparrow(t)$ in parametric resonance regime
(\ref{prob_periodic_switching}) does not contain high-frequency
oscillations at $\omega$. Moreover, the maximum probability is not
small anymore. This last feature is of great importance for the
possibility of an experimental investigation of the dynamical Lamb
effect.

We also wish to stress that the precise value of the maximum
excitation  probability, as given by Eq.
(\ref{prob_single_switching}), is rather relative. Namely, if we
take into account the next term of the perturbative expansion in Eq.
(\ref{solution}), the prefactor $2/\pi^2$ is going to change,
staying nevertheless independent on $g$. In addition, an admixture
of Rabi-like harmonic with higher frequency will appear, which
corresponds to oscillations between the two states with the number
of excitations equal to 4. Nevetheless, Eq.
(\ref{prob_periodic_switching}) yields correct qualitative
description for $w_\uparrow(t)$, as further evidenced by a
comparison with fully numerical results. A lack of a
quantitative agreement is due to the fact that our picture based on
the separation of fast and slow degrees of freedom is not so
accurate anymore because of a high-frequency modulation of $g$.

Next, we solve the time dependent Schr\" odinger equation
numerically.  We indeed see Rabi-like oscillations for both the
qubit excitation probability and the mean photon number, as shown by
Fig. 3. Frequencies of these oscillations are correctly described by
Eq. (\ref{prob_periodic_switching}). The estimated period of
$w_\uparrow(t)$ oscillations corresponds to the time scale
$\sqrt{2\pi}/ g $ appearing in the RWA Hamiltonian in the
two-excitation solution. We also found that maximum $w_\uparrow(t)$
and photon number $n_{ph}(t)$ are only slightly sensitive to the
amplitude of the coupling strength $g$ in the weak interaction
regime, which is again in a qualitative agreement with the simple
analytical treatment. At the same time, the value of maximum
$w_\uparrow(t)$, as predicted by Eq.
(\ref{prob_periodic_switching}), is nearly twice underestimated
compared to the exact (numerical) solution. Thus, both
$w_\uparrow(t)$ and $n_{ph}(t)$ demonstrate a universal behavior.
Namely, by rescaling the time, they can be cast in a universal form
within a few percent accuracy. This feature, which is predicted
qualitatively by Eq. (\ref{prob_periodic_switching}), is in a full
agreement with the results of our numerical computations.

\begin{figure}[h]
\center\includegraphics[width=0.5\linewidth]{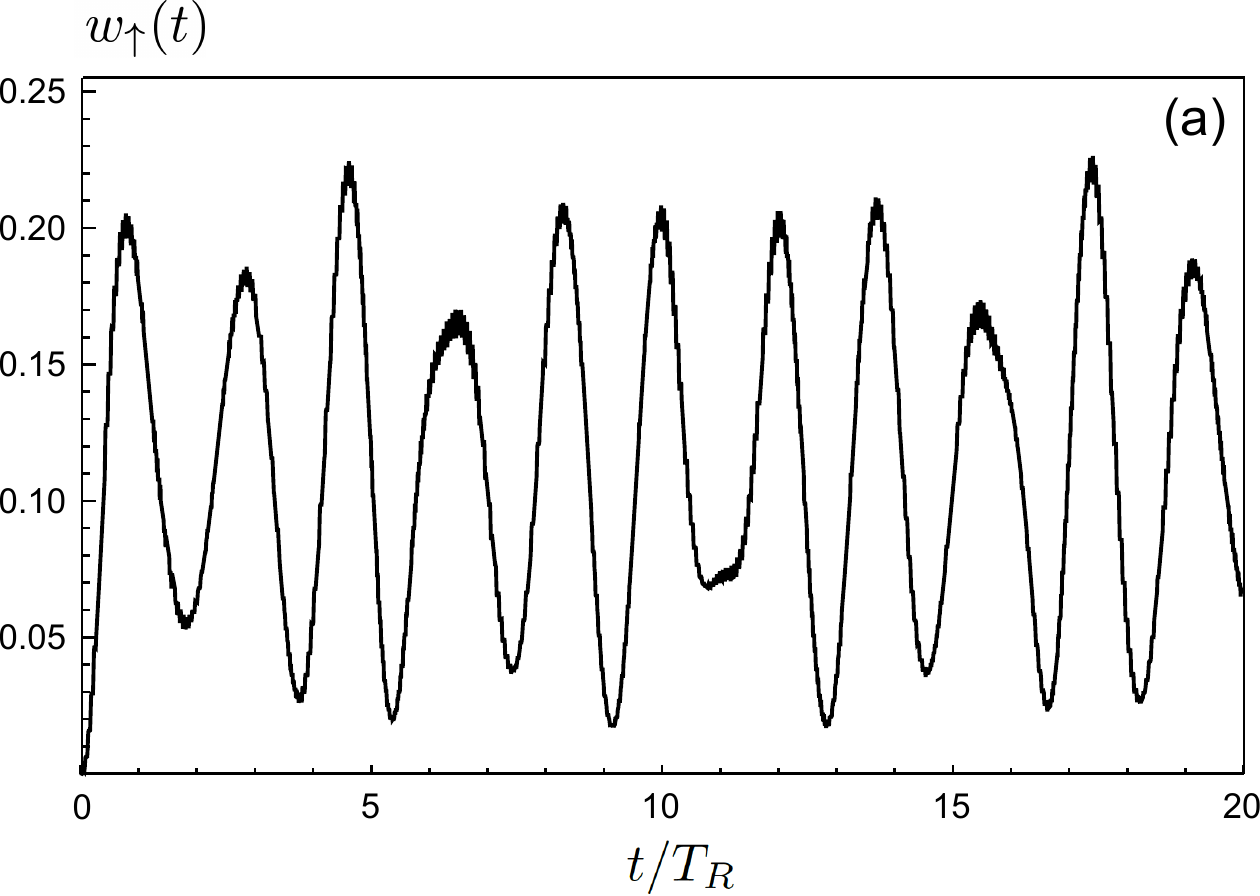}
\center\includegraphics[width=0.5\linewidth]{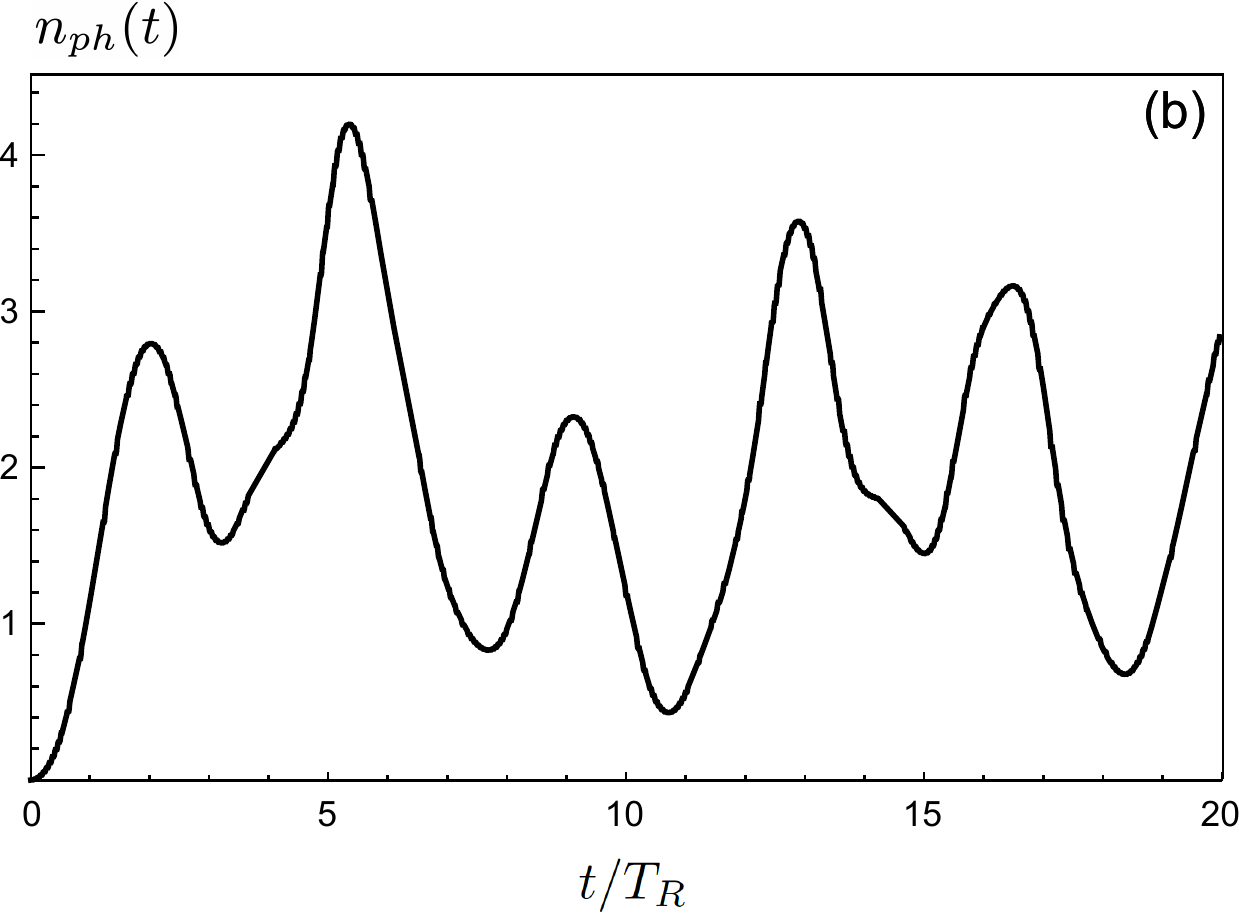}
\caption{ \label{parameteric_num_res} The qubit excitation probability $w_\uparrow(t)$ (a) and the
photon number $n_{ph}(t)$ (b) as functions of time after the beginning of parametric driving at $\omega/g=20$.}
\end{figure}

Recently, it was suggested theoretically that the Lamb  shift of a
qubit can be greatly enhanced via periodical driving of a qubit by
the external classical electromagnetic field \cite{Lambshift}. This
idea has some similarities with our idea of parametric pumping.
However, in contrast to Ref. \cite{Lambshift}, we do not need an
external classical field which directly affects a qubit and modifies
its Lamb shift. Instead we modulate only the qubit-resonator
coupling, while qubit is excited and photons appear as a result of
the interaction between a qubit and resonator.

Nevertheless, both ideas indicate that counter-rotating processes
in a weak coupling regime can be significantly enhanced by using
various types of a periodic driving. The same strategy was used to
increase a dynamical Casimir effect which finally has led to its
experimental observation \cite{DCE1,DCE2}. Thus, periodic driving,
in general, can be exploited to enhance various nonstationary cavity
QED effects.

\section{Correlation functions}

In addition to the qubit degrees of freedom, our system has also
photon ones. They can be characterized by various
correlation functions. For instance, one can use squeezing
parameters defined in a standard way as $\Delta p \equiv
\frac{1}{2}\sqrt{\langle (a-a^{\dagger })^2\rangle-\langle
a-a^{\dagger } \rangle^2}$ and $\Delta x \equiv
\frac{1}{2}\sqrt{\langle (a+a^{\dagger })^2\rangle-\langle
a+a^{\dagger } \rangle^2}$. For the stationary system, these two
parameters can be evaluated approximately using a perturbation
theory around RWA solution (see Appendix A). The wave function $\psi_0^{{{(1)}}}$ yields the following squeezing parameters
(in the first order by $g$)
\begin{equation}
\Delta p=\Delta x=\frac{1}{2}\left(
1-\frac{2g\alpha_{2}^{(-)}\beta_{2}^{(-)}}{\epsilon_{2,\psi}}
\right). \label{sq_stationary}
\end{equation}
It is easy to see that $\Delta p, \Delta x > 1/2$, since
$\alpha_{2}^{(-)}\beta_{2}^{(-)}<0$, which implies that this state
is not squeezed.

\begin{figure}[h]
\center\includegraphics[width=0.5\linewidth]{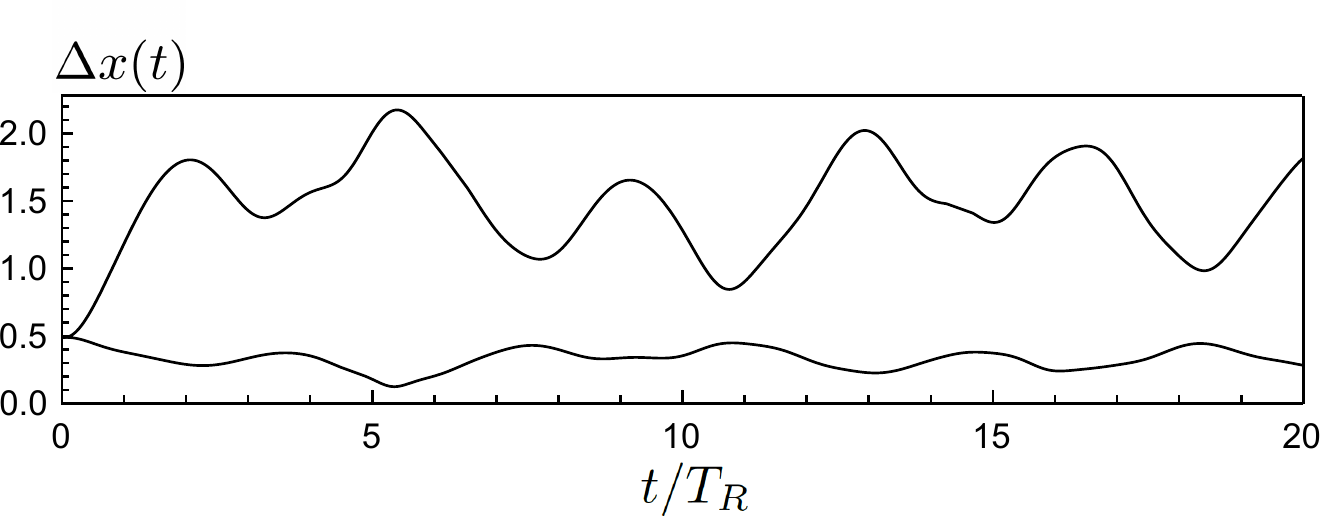}
\caption{ \label{squeezing_fig} Numerical result for the squeezing
parameter $\Delta x(t)$ at $\omega/g=20$  within
$20$ Rabi periods after the beginning of the parametric driving. Only lower and upper
envelope curves are shown, while $\Delta x(t)$ experiences fast oscillations between them.}
\end{figure}

Let us now consider squeezing in the case of a parametric modulation
of the qubit-resonator coupling constant. An accuracy of the
analytical solution in the first order of perturbation theory is
not sufficient for the quantitative analysis. We therefore
restrict ourselves to the numerical solution of the time dependent
Schr\"odinger equation taking into account nearly $30$ photon
states. Numerical solution shows the realization of the squeezed
state under parametric driving of a particular form
$g(t)=g\theta(\cos(2\omega t))$. We observe non-periodic
oscillations of the squeezing parameter $\Delta x$. Figure 4 shows
the lower and upper envelope curves for the
evolution of $\Delta x$ within first $20$ Rabi periods after the
switching on the driving $g(t)$ for $\omega/g=20$. These envelope curves experience
slow oscillations, while fast oscillations of $\Delta x$
occur between them. The fast oscillations are not shown in Fig. 4.

The results again reveal the universality of the envelope curves oscillations,
regardless of the ratio $\omega/g$. The minimum value of $\Delta x$
at particular moments of time, according to Fig. 4, is $\approx
1/4$, which indicates a sufficiently strong squeezing of a generated
photon state. By comparison Fig. 4 and Fig. 3, we see that oscillations of
$\Delta x$ are correlated with oscillations of $n_{ph}(t)$.

\begin{figure}[h]
\center\includegraphics[width=0.5\linewidth]{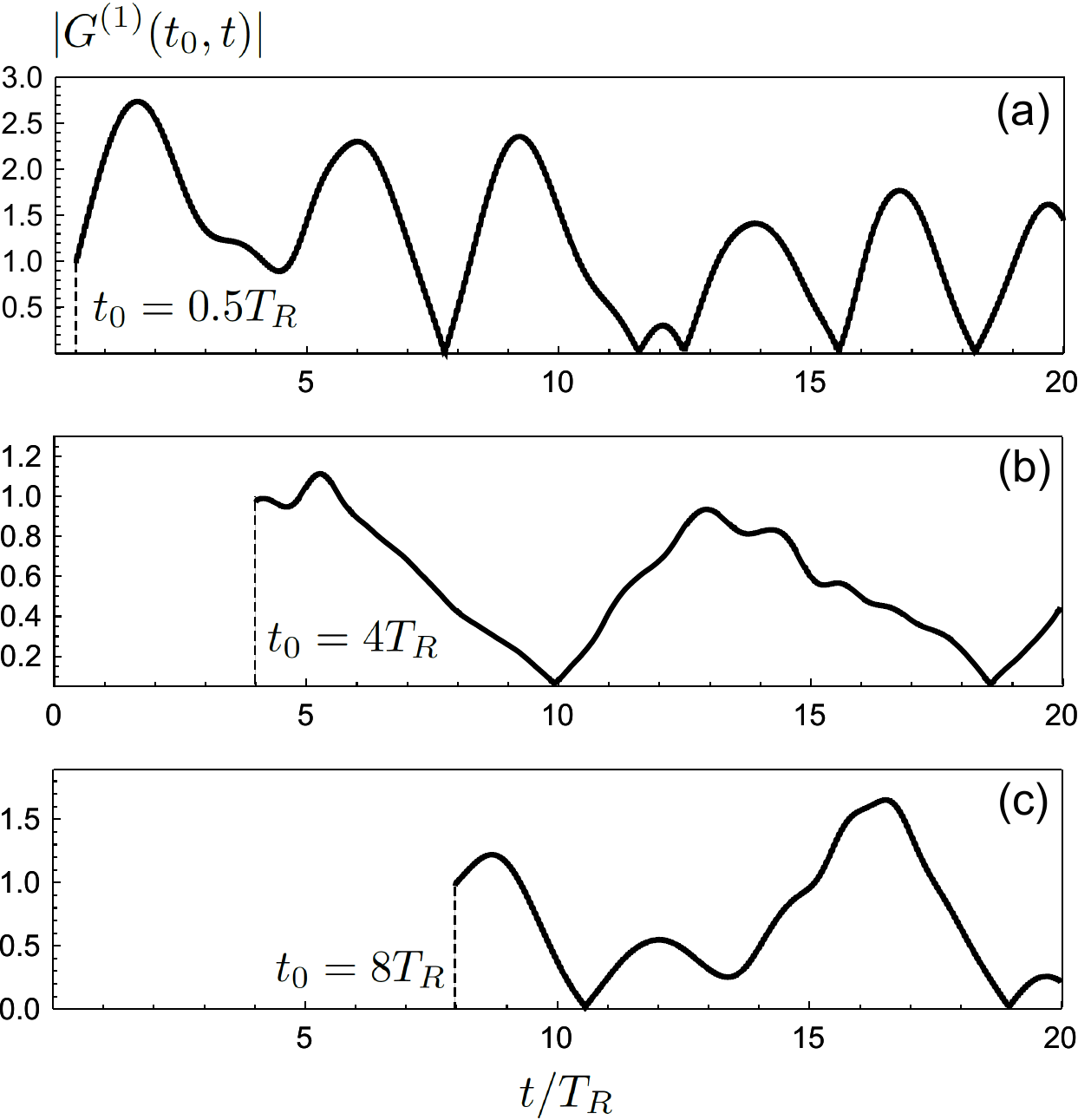}
\caption{ \label{aa} Numerical result for photon correlation
functions $G^{(1)}(t_0,t)$ at $t_0=0.5 T_R$ (a), $4T_R$ (b), $8T_R$
(c) at $\omega/g=20$ within $20$ Rabi periods after the starting of
the parametric driving.}
\end{figure}

\begin{figure}[h]
\center\includegraphics[width=0.5\linewidth]{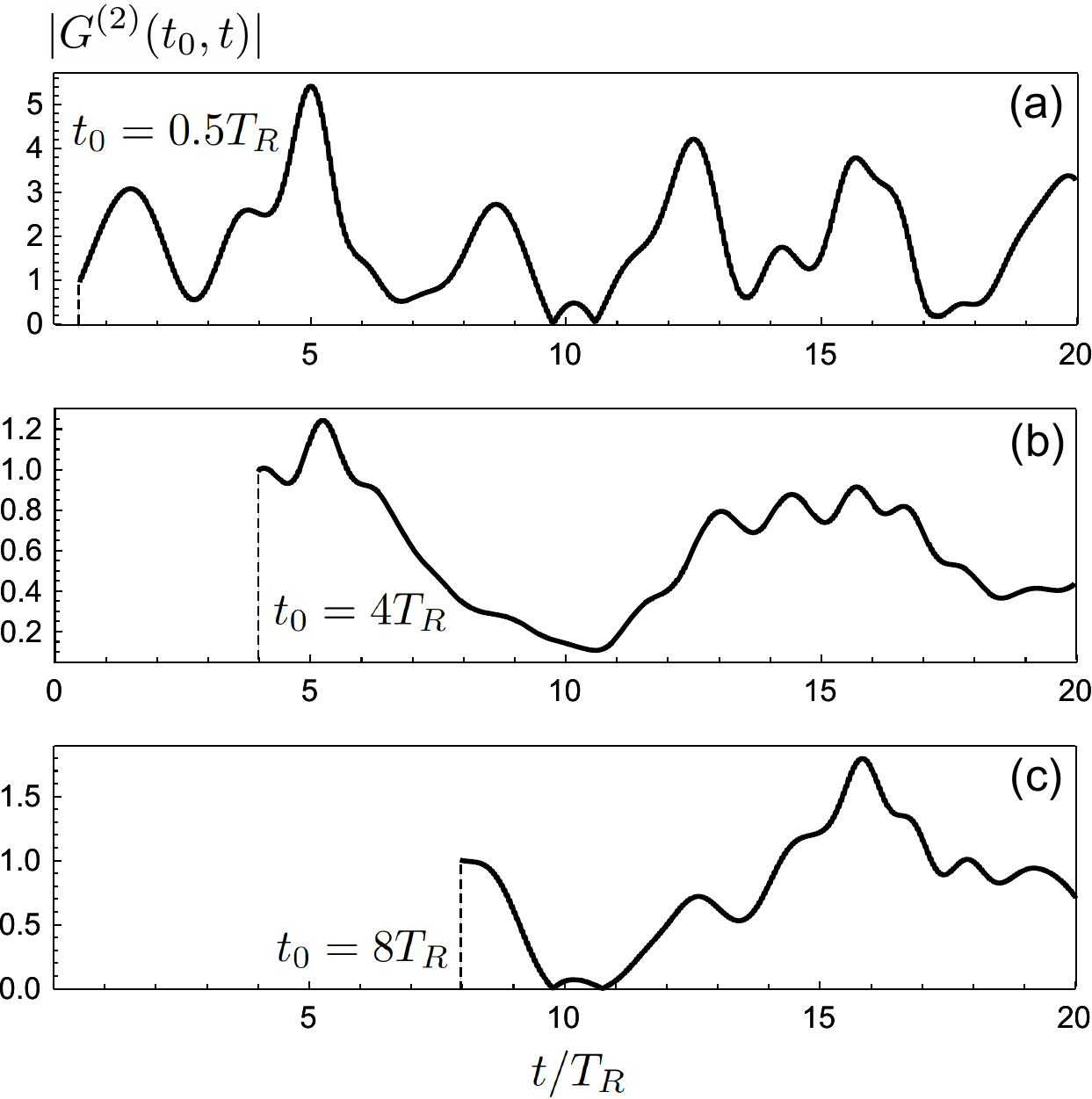}
\caption{ \label{aaaa}  Numerical result for photon correlation
functions $G^{(2)}(t_0,t)$ at $t_0=0.5 T_R$ (a), $4T_R$ (b), and
$8T_R$ (c) at $\omega/g=20$ within $20$ Rabi periods after the
starting of the parametric driving. }
\end{figure}

The first-order and second-order correlation functions can be expressed
through the field operators in a Heisenberg picture as
\begin{eqnarray}
& G^{(1)}(t_0,t)=\frac{\langle a^{\dagger }(t_0)a(t) \rangle}{\langle
a^{\dagger }(t_0)a(t_0) \rangle}, \notag \\
& G^{(2)}(t_0,t)=\frac{\langle a^{\dagger }(t_0)a(t_0) a^{\dagger
}(t)a(t)\rangle}{\langle a^{\dagger }(t_0)a(t_0)a^{\dagger }(t_0)a(t_0)
\rangle}.
\end{eqnarray}
These two quantities are calculated using a numerical solution of
the time-dependent Schr\"{o}dinger equation. Typical results are
presented  in Figs. 5 and 6. Both correlation functions demonstrate
non-periodic oscillations with slow decay as functions of $t-t_0$,
where $t_0$ corresponds to the starting point and the correlation is
probed at $t>t_0$. By comparing these two figures with Fig 3, it is
easy to see that the correlation functions at small $t_0$ generally
follow the field revivals as a function of $t$, while oscillations
are not so pronounced at larger $t_0$. The existence of slow
oscillations is clearly linked to the interplay between the qubit
and photon field, which is accompanied by the qubit excitation due
to the dynamical Lamb effect.

We also introduce first-order and second-order correlation functions
for qubit degrees of freedom (qubit excitation number) as
\begin{eqnarray}
& G_{q}^{(1)}(t_0,t)=\frac{\langle \sigma _{+}(t_0) \sigma _{-}(t)
\rangle}{\langle
\sigma _{+}(t_0) \sigma _{-}(t_0) \rangle}, \notag \\
& G_{q}^{(2)}(t_0,t)=\frac{\langle \sigma _{+}(t_0) \sigma _{-}(t_0)
\sigma _{+}(t) \sigma _{-}(t) \rangle}{\langle \sigma _{+}(t_0)
\sigma _{-}(t_0) \sigma _{+}(t_0) \sigma _{-}(t_0) \rangle},
\end{eqnarray}
which are again calculated numerically. The results are shown in
Figs. 7 and 8 (for the same set of parameters as for Figs. 5 and 6).
The behavior of these correlation functions generally resembles the
behavior of similar quantities for the photon field. The dynamics of
$G_{q}^{(1)}$ and $G_{q}^{(2)}$ follows the dynamics of
$w_\uparrow(t)$.

\begin{figure}[h]
\center\includegraphics[width=0.5\linewidth]{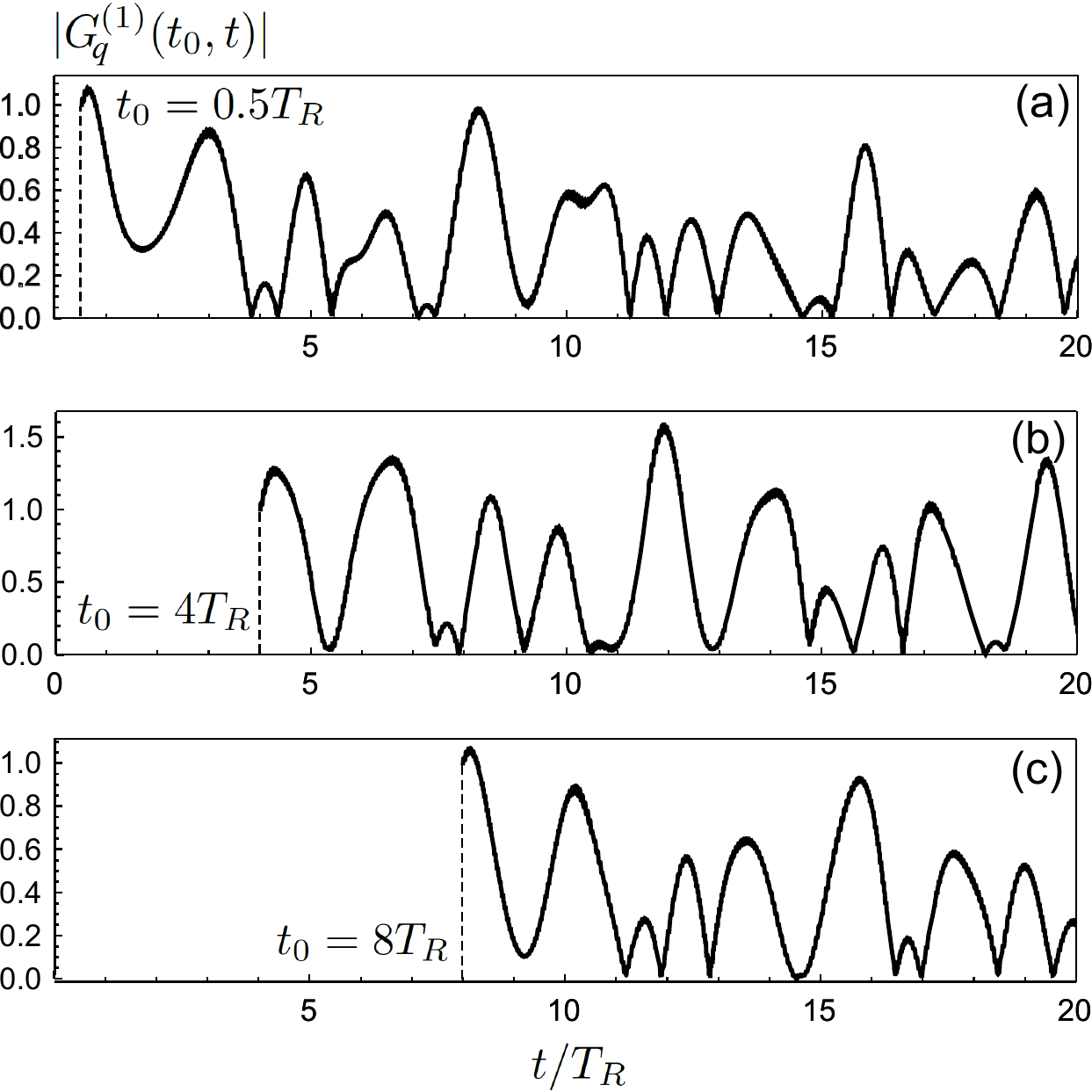}
\caption{ \label{qqqq} Numerical result for qubit correlation
functions $G_{q}^{(1)}(t_0,t)$ at $t_0=0.5 T_R$ (a), $4T_R$ (b),
$8T_R$ (c) at $\omega/g=20$ within $20$ Rabi periods after the
starting of the parametric driving.}
\end{figure}

\begin{figure}[h]
\center\includegraphics[width=0.5\linewidth]{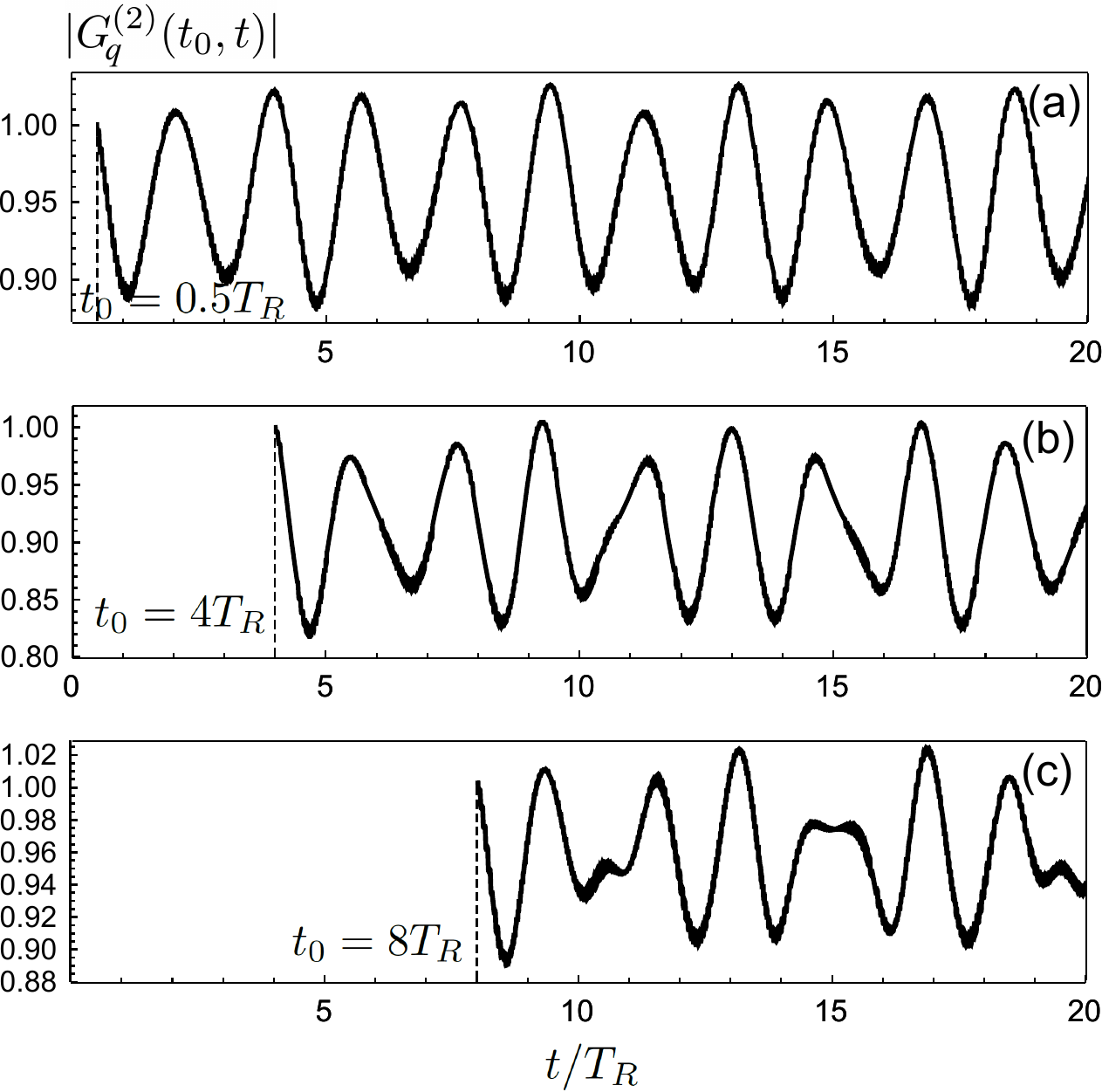}
\caption{ \label{qq}  Numerical result for qubit correlation
functions $G_{q}^{(2)}(t_0,t)$ at $t_0=0.5 T_R$ (a), $4T_R$ (b), and
$8T_R$ (c) at $\omega/g=20$ within $20$ Rabi periods after the
starting of the parametric driving. }
\end{figure}

However, the amplitude of oscillations of $G_{q}^{(2)}$ is rather
small, while this quantity does not experience a decay as a function
of $t-t_0$. This is related to the two-level character of qubit
energy spectrum which has to be contrasted with the unbounded photon
spectrum. Moreover, for the considered range of parameters, photons
occupy several levels, while qubit mostly stays in its ground state.
This latter feature leads to only small oscillations of
$G_{q}^{(2)}$. Nevertheless, slow oscillations of correlation
functions for the photon field, which can be probed in experiments,
are definitely related to slow oscillations of correlation functions
for qubit degrees of freedom, as well as to the qubit excitation
probability.

Thus, slow Rabi-like oscillations of various characteristics of a
photon field can be used in experiments as a signature of the
dynamical Lamb effect in the systems with tunable photon-qubit
coupling. Indeed, if the resonator was initially empty and qubit was
uncoupled from it, no such oscillations can exist, of course. If a
coupling energy is modulated, qubit can be excited leading to photon
creation via counter-rotating processes, which can be seen in
experiments. In contrast to the dynamical Casimir effect, photons
occupy not only even states.

\section{Conclusions}
 \label{summary}

We suggested that a coupled superconducting qubit-resonator system
can be used for the observation of the dynamical Lamb effect. This
effect was initially introduced for the natural atom in a varying
cavity \cite{Lozovik1}. It leads to the probability for the atom to
be excited solely due to the nonadiabatic modulation of atomic
levels Lamb shift. Although this effect was initially predicted for
natural atoms, it is quite problematic to observe it in experiments
with such atoms, since this effect must be carefully separated from
other mechanisms of atom excitation. To this end, it was proposed in
Ref. \cite{Lozovik1} to pass a single atom through the resonator
consisting of two cameras with different diameters. When passing the
cameras, an atom can be excited and a photon can be generated. We
suggest a nontrivial link between this scheme and a system with a
single superconducting qubit (macroscopic artificial atom) \emph{at
rest} integrated with the resonator in such a way that the coupling
can be tuned dynamically. By switching on this coupling
nonadiabatically, one can excite a qubit and generate a photon, even
if the resonator was initially unoccupied by photons. Such a
dynamical coupling between the qubit (transmon) and the resonator is
already achievable by utilizing an auxiliary SQUID
\cite{tunable_qubit}. Thus, we suggest that this scheme can be used
for the experimental realization of the dynamical Lamb effect.

We presented a theoretical model for the description of the system
dynamics upon a modulation of the coupling energy and for the predictions of
a method to increase the effect. Our
treatment is based on the dynamical Rabi model in which we take
into account counter-rotating terms, since they are responsible
for the dynamical Lamb effect. The dynamical problem was solved both
analytically using a perturbation theory around the RWA stationary
solution and numerically by integrating time dependent
Schr\"{o}dinger equation.

We found that a single switching of the coupling constant results
only in a low probability for the qubit to be excited when
considering weak coupling regime, as well as a full resonance
between the qubit and the cavity single mode. After the switching
on, this probability experiences slow Rabi-like oscillations. We
then suggested a much more efficient method to enhance the effect
which is a periodic nonadiabatical switching of the coupling energy.
We also studied a statistics of photon states generated upon the
coupling constant modulation. We found that these states can
experience a strong squeezing.

\begin{acknowledgments}
The authors acknowledge useful comments by A. V. Ustinov, A. L. Rakhmanov, V. P. Yakovlev, and
L. V. Bork. This work was supported by RFBR (project no. 15-02-02128). W. V. P. acknowledges the support from the Russian Science Support Foundation. Yu. E. L. is supported by Program of Basic Research of HSE.
\end{acknowledgments}

\appendix

\section{Perturbation theory around RWA solution}

The stationary Rabi Hamiltonian can be split as
 \begin{equation}
H=H_{RWA}+V_2.
  \label{split1}
\end{equation}
The first part $H_{RWA}$ is the Hamiltonian in the rotating wave approximation,
 \begin{equation}
H_{RWA}=\omega a^{\dagger }a + \frac{1}{2} \epsilon (1+\sigma_{3})+V_{1},
  \label{RWA}
\end{equation}
 which is known to be exactly integrable.

The set of the ''dressed'' eigenstates and corresponding eigenenergies of $H_{RWA}$ is given by the following expressions
\begin{eqnarray}
\psi_n^{(0)} =\alpha_{n}^{(-)}|  n \downarrow\rangle + \beta_{n}^{(-)} |  n-1 \uparrow\rangle, \quad \epsilon_{n,\psi}^{(0)} \label{psi_n}\\
\chi_n^{(0)} = \alpha_{n}^{(+)}|  n \downarrow\rangle + \beta_{n}^{(+)} |  n-1 \uparrow\rangle, \quad \epsilon_{n,\chi}^{(0)}\nonumber
\end{eqnarray}
where $ |  n \downarrow\rangle$ and $|  n-1 \uparrow\rangle$ form a set of ''bare'' states of the decoupled qubit and photons. Coefficients in (\ref{psi_n}) read as ($n\geq 1$):
\begin{equation}
\alpha_{n}^{(\pm)}=\frac{g\sqrt{n}}{\sqrt{\frac{\Delta^2}{2}+2g^2 n \pm \Delta\sqrt{\frac{\Delta^2}{4}+g^2 n}}},    \label{abdg}
\end{equation}
$$ \beta_{n}^{(\pm)}=\frac{ \frac{\Delta}{2} \pm \sqrt{\frac{\Delta^2}{4}+g^2 n} }{\sqrt{\frac{\Delta^2}{2}+2g^2 n -\Delta\sqrt{\frac{\Delta^2}{4}+g^2 n}}}.$$
A bare ground state $ |  0 \downarrow\rangle$ is not affected by
$V_1$ and remains the same as at $g=0$. Thus, there is only one
state corresponding to $n=0$, which is $\psi_0^{(0)}$:
$$
 \psi_0^{(0)} =|  0 \downarrow\rangle.
$$

The eigenenergies of $H_{RWA}$ are
\begin{equation}
\epsilon_{n,\psi}=\omega n +\frac{\Delta}{2}-\sqrt{\frac{\Delta^2}{4}+g^2 n}, \quad n\geq 0,  \label{energy}
\end{equation}
$$
\epsilon_{n,\chi}=\omega n +\frac{\Delta}{2}+\sqrt{\frac{\Delta^2}{4}+g^2 n}, \quad n\geq 1.
$$

Using ô standard perturbation theory, we find that the wave functions within the first order contribution in $V_2$ read ($n\geq 1$)
 \begin{widetext}
\begin{equation}
\psi_n^{(1)}= \psi_n^{(0)} + g\left( \frac{\alpha_{n-2}^{(-)}\beta_{n}^{(-)}\sqrt{n-1}}{\epsilon_{n,\psi}-\epsilon_{n-2,\psi}}  \psi_{n-2}^{(0)}+\frac{\beta_{n+2}^{(-)}\alpha_{n}^{(-)}\sqrt{n+1}}{\epsilon_{n,\psi}-\epsilon_{n+2,\psi}}  \psi_{n+2}^{(0)} + \frac{\alpha_{n-2}^{(+)}\beta_{n}^{(-)}\sqrt{n-1}}{\epsilon_{n,\psi}-\epsilon_{n-2,\chi}}  \chi_{n-2}^{(0)} + \frac{\beta_{n+2}^{(+)}\alpha_{n}^{(-)}\sqrt{n+1}}{\epsilon_{n,\psi}-\epsilon_{n+2,\chi}}  \chi_{n+2}^{(0)} \right). \label{wf1}
\end{equation}
\begin{equation}
\chi_n^{(1)}= \chi_n^{(0)} + g\left( \frac{\alpha_{n-2}^{(+)}\beta_{n}^{(+)}\sqrt{n-1}}{\epsilon_{n,\chi}-\epsilon_{n-2,\chi}}  \chi_{n-2}^{(0)}+\frac{\beta_{n+2}^{(+)}\alpha_{n}^{(+)}\sqrt{n+1}}{\epsilon_{n,\chi}-\epsilon_{n+2,\chi}}  \chi_{n+2}^{(0)} + \frac{\alpha_{n-2}^{(-)}\beta_{n}^{(+)}\sqrt{n-1}}{\epsilon_{n,\chi}-\epsilon_{n-2,\psi}}  \psi_{n-2}^{(0)} + \frac{\beta_{n+2}^{(-)}\alpha_{n}^{(+)}\sqrt{n+1}}{\epsilon_{n,\chi}-\epsilon_{n+2,\psi}}  \psi_{n+2}^{(0)}\right). \label{wf2}
\end{equation}
\end{widetext}

For the modified ground state ($n=0$), we obtain
\begin{equation}
\psi_0^{(1)}= |0 \downarrow\rangle - \frac{g\beta_{2}^{(-)}}{\epsilon_{2,\psi}}(\alpha_{2}^{(-)} |2 \downarrow\rangle +\beta_{2}^{(-)}|1 \uparrow\rangle).
\end{equation}
%The superposition of the ''bare'' ground state with the doubly excited ones in the $|  \psi_0^{(1)}\rangle$ comes from  the $b^+b(a^+)^2$ from the $V_2$.

Using the above strategy, one can also find corrections in $V_2$ to the eigenstates of $H_{RWA}$.
For instance, a first non-zero contribution to the ground state energy is determined by the second order correction and is given by the negative value
\begin{equation}
\epsilon_{0,\psi}^{(2)}  %2 \sum\limits_k \left( \frac{|V_{2\psi}^{0k}|^2}{\epsilon_{0,\psi}-\epsilon_{k,\psi}}+\frac{|V_{2\chi}^{0k}|^2}{\epsilon_{0,\chi}-\epsilon_{k,\chi}}\right)
= - \frac{g^2|\beta_{2}^{(-)}|^2}{\epsilon_{2,\chi}}= \frac{ -4g^2(1 - \Delta/\sqrt{\Delta^2 +8g^2 })}{3\omega+\epsilon - \sqrt{\Delta^2 +8g^2 } }. \label{lamb_res}
%, \quad \Delta=\omega-\epsilon.
\end{equation}
This is the Lamb shift value, where all the processes, generated by the rotating wave term $V_1$ in the Hamiltonian, are included in zero-order approximation. This scheme of the expansion with respect to the operator $V_2$ around $H_{RWA}$ regularizes the perturbation theory against $1/\Delta$ divergencies. It is applicable in the opposite resonant case $|\Delta|\ll g$, where the perturbative result of Ref. \cite{Lozovik1} is not adequate due to divergent corrections $\sim g/ \Delta$ to the wave functions and eigen energies. The divergencies originate from the degeneracy of the self-mode and qubit energy.

It is of importance to note that Eq. (\ref{lamb_res}) does not match
the off-resonant result (\ref{lamb_off-res}) that means that this
$V_2$-expansion is most reliable in the resonant limit $\Delta\ll
g$, despite that the detuning $\Delta$ is formally not restricted in
this approach. In order to recover the off-resonant  result
(\ref{lamb_off-res}) we have to consider contributions from higher
orders of $V_2$.

\end{document}